\documentclass[runningheads]{llncs}
\usepackage{graphicx}
\usepackage{multirow}
\usepackage{hyperref}
\usepackage{booktabs}
\usepackage{bbding}
\usepackage{amsfonts}

\usepackage{algorithm,algcompatible}

\algnewcommand\INPUT{\item[\textbf{Input:}]}%
\algnewcommand\OUTPUT{\item[\textbf{Output:}]}%
\renewcommand{\COMMENT}[2][.6\linewidth]{\leavevmode\hfill\makebox[#1][l]{\#~#2}}

\begin{document}
\title{Joint Motion Correction and Super Resolution for Cardiac Segmentation via Latent Optimisation}

\titlerunning{Joint Motion Correction and Super Resolution }

\author{Shuo Wang\inst{1,2,3}(\Envelope), 
Chen Qin \inst{4,5}, 
Nicolò Savioli \inst{4,6}, 
\\
Chen Chen \inst{4}, 
Declan O'Regan \inst{6}, 
Stuart Cook \inst{7}, 
\\
Yike Guo\inst{3}, 
Daniel Rueckert\inst{4}, 
and Wenjia Bai\inst{3,8}} 

\authorrunning{S. Wang et al.}

\institute{
Digital Medical Research Center, Fudan University, China
\and 
Shanghai Key Laboratory of MICCAI, China
\and Data Science Institute, Imperial College London, UK 
\and Department of Computing, Imperial College London, UK 
\and Institute for Digital Communications, University of Edinburgh, UK
\and
MRC London Institute of Medical Sciences, Imperial College London, UK
\and
National Heart Research Institute, Singapore
\and Department of Brain Sciences, Imperial College London, UK
\email{shuowang@fudan.edu.cn}}%

\authorrunning{S. Wang et al.}
\maketitle              
\begin{abstract}
In cardiac magnetic resonance (CMR) imaging, a 3D high-resolution segmentation of the heart is essential for detailed description of its anatomical structures. However, due to the limit of acquisition duration and respiratory/cardiac motion, stacks of multi-slice 2D images are acquired in clinical routine. The segmentation of these images provides a low-resolution representation of cardiac anatomy, which may contain artefacts caused by motion. Here we propose a novel latent optimisation framework that jointly performs motion correction and super resolution for cardiac image segmentations. Given a low-resolution segmentation as input, the framework accounts for inter-slice motion in cardiac MR imaging and super-resolves the input into a high-resolution segmentation consistent with input. A multi-view loss is incorporated to leverage information from both short-axis view and long-axis view of cardiac imaging. To solve the inverse problem, iterative optimisation is performed in a latent space, which ensures the anatomical plausibility. This alleviates the need of paired low-resolution and high-resolution images for supervised learning. Experiments on two cardiac MR datasets show that the proposed framework achieves high performance, comparable to state-of-the-art super-resolution approaches and with better cross-domain generalisability and anatomical plausibility. The codes are available at \url{https://github.com/shuowang26/SRHeart}.

\end{abstract}
%

\section{Introduction}
\begin{figure}
\begin{center}
\includegraphics[width=\textwidth]{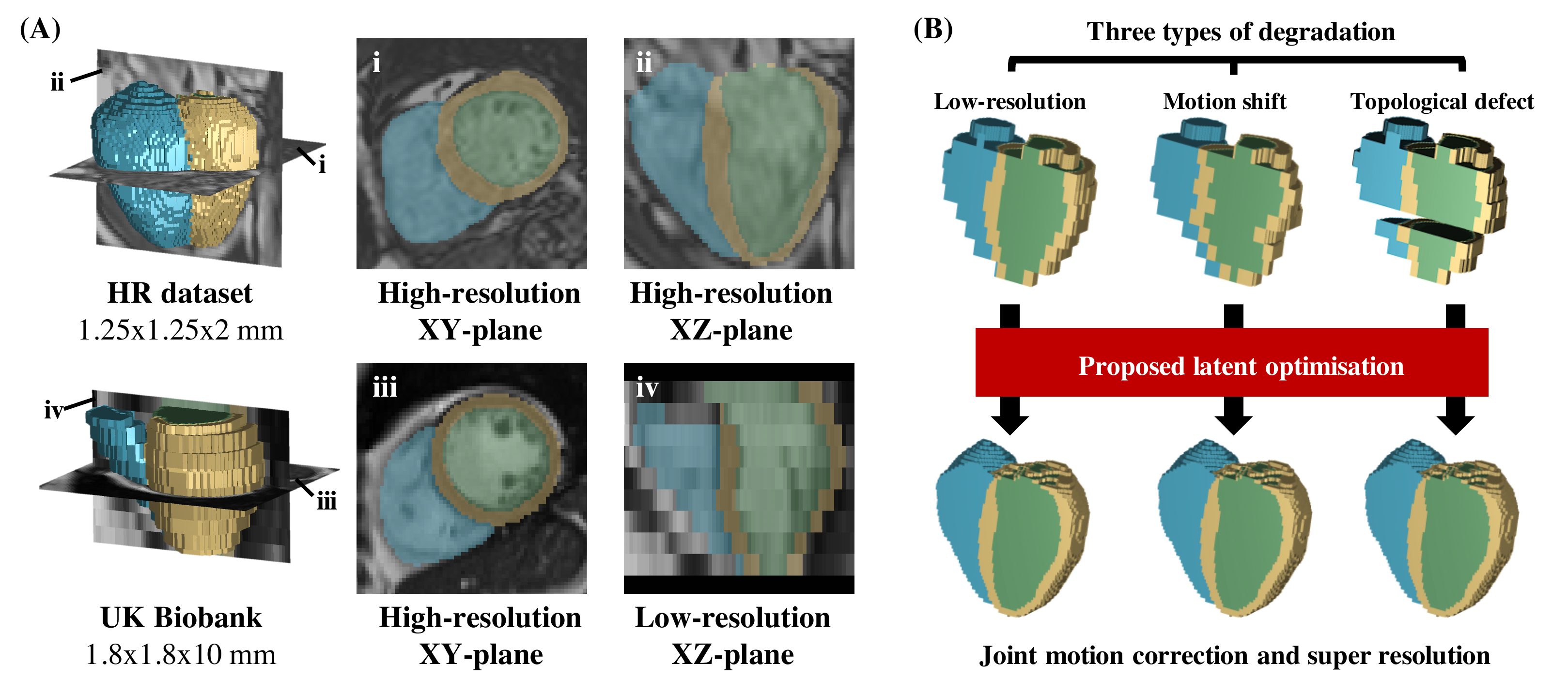}
\end{center}
   \caption{(A) Illustration of the difference between a high-resolution volume acquired by a research protocol (HR dataset, top) and a low-resolution stack of slices acquired by a standard clinical protocol (UK Biobank, bottom). The UK Biobank data is of low resolution along the Z-axis (subfigure iv: XZ-plane). (B) Illustration of typical degradations in cardiac segmentation.}
\label{Fig1}
\end{figure}

In cardiac imaging, a high-resolution geometric representation of the heart is desired for accurate assessment of its anatomical structure and function. However, this is not easily available in clinical practice or research. For example, cardiac magnetic resonance (CMR) imaging is the current gold standard imaging modality of the heart. Although high-resolution 3D volumes may be acquired with a research acquisition protocol (Figure \ref{Fig1}A, top row), it is not applicable to the clinical routine, as the research protocol requires a long breath-hold 
that is not feasible for patients with severe cardiac diseases. Instead, a stack of 2D short-axis slices are usually acquired at multiple breath-holds (Figure \ref{Fig1}A, bottom row). Segmentations built from these 2D slices suffer from three types of degradations: anisotropic low-resolution, motion shifts and potential topological defects (Figure \ref{Fig1}B). To address these issues, we propose a novel latent optimisation framework for joint motion correction and super resolution of 3D cardiac image segmentations.

\noindent\textbf{Related work:} There is a large corpus of works on image super-resolution, including both deep learning-based methods \cite{Wang2020} and regularised optimisation methods \cite{Yue2016}. Here, we focus on reviewing methods based on convolutional neural networks (CNNs) and related to cardiac imaging enhancement.

\noindent\textbf{Supervised learning for super-resolution:}
Since the first attempt using CNN for single image super-resolution (SR) \cite{dong2015image}, deep learning-based SR algorithms have successfully transformed the state-of-the-art performance \cite{he2016deep,qiu2019embedded,tong2017image,zhang2018image,bulat2018learn}. In general, these SR networks are trained in a supervised way where pairs of low-resolution (LR) and high-resolution (HR) images are provided. The network is trained to approximate the inverse process of image degradation. However, a large dataset of real paired LR-HR images is often not available but to be simulated, imposing over-fitting risks. Another drawback is their potential failure in cross-domain applications. A well-trained neural network may generate implausible results if applied on a dataset from an unseen domain.

\noindent\textbf{Latent space exploration:}
Instead of approximating the LR$\rightarrow$HR mapping through supervised learning, the inverse problem also can be solved via optimisation. Given an LR input, optimisation can be performed in the HR space till the best-fit HR image is found. However, optimisation in the high-dimensional HR image space is challenging and manifold learning has been introduced to shrink the dimensionality \cite{ulyanov2018deep,law2006incremental,zhu2018image}. Menon et al. proposed the PULSE framework to learn the manifold of HR face images \cite{Menon_2020_CVPR}. This method does not require paired LR-HR images for supervised training but explores the latent space. In parallel, the concept of latent space exploration has been introduced in medical imaging applications including reconstruction \cite{Tezcan2018}, quality control \cite{wang2020deep} and refinement \cite{painchaud2020cardiac}. These works share the view that images or segmentations are distributed on a low-dimensional manifold which can be parameterised by latent codes. 

\noindent\textbf{Cardiac imaging enhancement:}
The main aim of cardiac super-resolution is to improve the resolution along the Z-axis (Figure \ref{Fig1}A). Many studies have investigated the cardiac image enhancement using either traditional methods or deep learning-based methods \cite{bhatia2014super,oktay2016multi}. A particular challenge in CMR imaging is the inter-slice misalignment caused by respiratory motion. For a stacked CMR volume, there exists a potential shift between adjacent 2D XY-plane image, which needs to be corrected for \cite{avendi2016combined,avendi2016combined,tarroni2018learning,odille2015motion}. Besides images, machine learning algorithms have also been proposed for enhancing the segmentations \cite{oktay2017anatomically,duan2019automatic,larrazabal2020post}. In this paper, we focus on the enhancement of cardiac segmentations. Working with segmentations provides us with the convenience to constrain the anatomical geometry and allows better generalisability across different domains.

\noindent\textbf{Contributions:} There are three major contributions of this work. Firstly, we propose a novel framework for joint motion correction and super resolution, which alleviates the need of paired samples and achieves competitive results. Secondly, the proposed approach analytically integrates multiple views without retraining the network. Finally, since the framework is based on segmentations, it is agnostic to imaging modalities, devices or contrasts.

\section{Methods}

\subsection{Problem formulation and preliminaries}
A typical CMR segmentation is a 3D volume that consists of a stack of 2D segmentations for each short-axis slices. Denote the segmentation by $S = \{y_i\}_{i \in N}$, where $y_i \in T = \{1, 2, ... C\}$ denotes the segmentation at $i$-th voxel, $C$ denotes the number of tissue types, $N$ denotes the total number of voxels, $N=D \times H \times W $, with $D$, $H$, $W$ being the depth, height and width. The low-resolution segmentation is denoted by $S_{LR} \in T^{D_{LR} \times H_{LR} \times W_{LR}}$. The corresponding high-resolution segmentation is $S_{HR} \in T^{D_{HR} \times H_{HR} \times W_{HR}}$. The degradation of cardiac segmentation, caused by low-resolution acquisition and motion, is formulated as:
\begin{equation}
S_{LR} = F(S_{HR}) = M_{d}(\downarrow_s {S_{HR}}) +\epsilon
\label{Eq1}
\end{equation}
where $F$ represents the degradation process of cardiac segmentation, including three components: a down-sampling operator $\downarrow_s$, an inter-slice motion operator $M_d$ and an error term $\epsilon$. $s=[D_{HR}/D_{LR}, H_{HR}/H_{LR}, W_{HR}/W_{LR}]$ is the down-scaling factor. The motion operator $M_d$ is parameterised by $d=[(d_H^1, d_W^1), (d_H^2, d_W^2), ..., (d_H^D, d_W^D)]$, which describes the displacements of 2D slices. $\epsilon \in T^{D_{LR} \times H_{LR} \times W_{LR}}$ represents voxel-wise segmentation errors. Equation (\ref{Eq1}) describes the HR$\rightarrow$LR degradation process with parameters $s$, $d$ and $\epsilon$. The task of joint motion correction and super resolution is that, given a degraded segmentation $S_{LR}$, restore a super-resolved segmentation $S_{SR}$ as close to the ground-truth $S_{HR}$ as possible. We briefly describe two commonly used techniques, direct interpolation and supervised learning-based interpolation, before presenting our proposed framework.

\noindent \textbf{Direct interpolation.} The simplest way is to apply a nearest neighbour (NN) up-sampling operation on the low-resolution segmentation, i.e. $S_{NN}=\uparrow_sS_{LR}$. Alternatively, shape-based interpolation (SBI) \cite{raya1990shape} can be used, which accounts for the shape context.

\noindent \textbf{Learning-based interpolation.} Similar to image super-resolution, we can utilise neural networks to approximate the inverse process $F^{-1}$ from low-resolution to high-resolution segmentations, i.e., LR$\rightarrow$HR mapping. Paired LR and HR samples are needed to train the networks.

\subsection{Proposed latent optimisation framework} 
\begin{figure}[h]
\begin{center}
\includegraphics[width=1\textwidth]{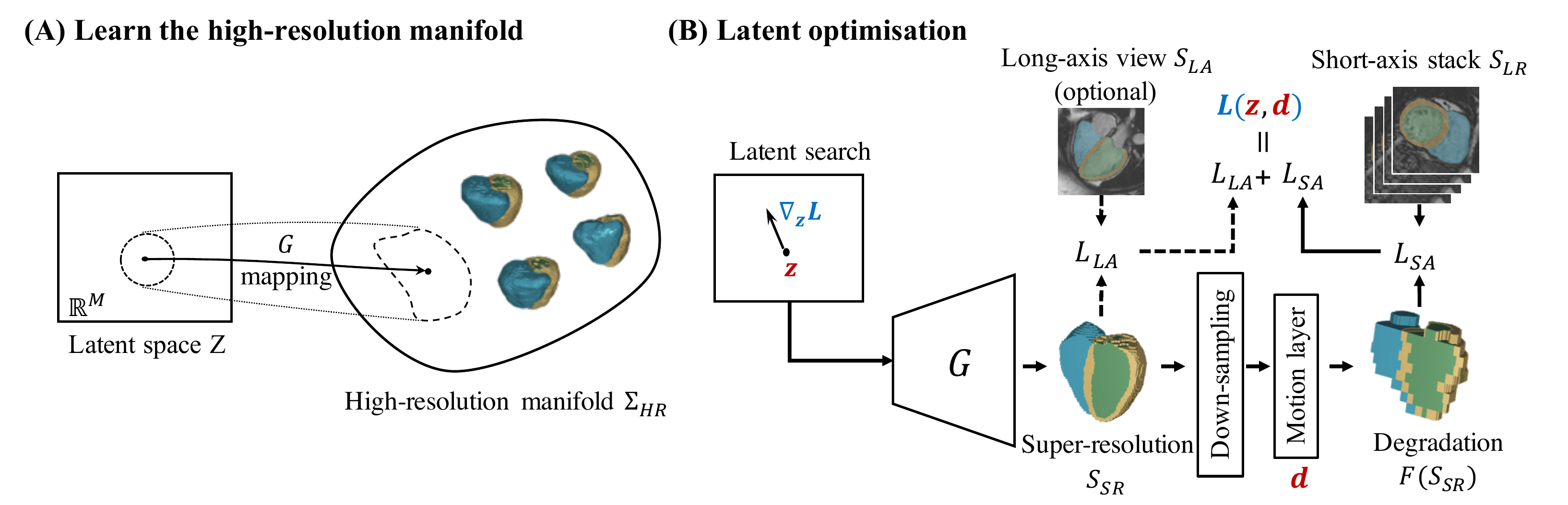}
\end{center}
   \caption{The latent optimisation framework for joint motion correction and super resolution. (A) A manifold of plausible high-resolution cardiac segmentations is learnt with a generative model $G$ that maps a low-dimensional space $Z$ to the manifold of high-resolution segmentation space $\Sigma_{HR}$. (B) Given a low-resolution segmentation $S_{LR}$, we search in the latent space via gradient descent to find a plausible high-resolution segmentation $S_{SR}$. The latent variable $z$ and motion displacement $d$ (annotated in red) are optimised to minimise the loss defined on short-axis and long-axis views.}
\label{Fig2}
\end{figure}

The proposed framework is composed of two stages. Firstly, we utilise a deep generative model to learn the latent space of plausible high-resolution segmentations (Figure \ref{Fig2}A). Then we perform iterative optimisation for the latent variable $z$ and the motion parameter $d$, named as `latent optimisation' (LO). The search is guided by the gradient backpropagated from a consistency loss comparing the degraded high-resolution segmentation by the model with the input low-resolution segmentation (Figure \ref{Fig2}B). 

\noindent\textbf{Anatomical prior.}
A key component of the proposed framework is the generative model $G$ which learns the manifold  $\Sigma_{HR} $ of plausible high-resolution cardiac segmentations (Figure \ref{Fig2}A), such that:
\begin{equation}
G_{\theta}(z): \mathbb{R}^{M} \ni z \mapsto S=G_{\theta}(z) \in \Sigma_{HR}
\label{Eq2}
\end{equation}
where $z$ denotes the latent representation in a low-dimensional space $\mathbb{R}^{M} (M\ll N)$ and $\theta$ denotes the parameters of the generative model. The generative model can be implemented with different architectures, e.g. VAEs and GANs. With a well-trained generative model, we expect a plausible segmentation via sampling in the latent space, imposing the anatomical prior for latent optimisation.

\noindent\textbf{Manifold traversing.} Given a degraded segmentation $S_{LR}$, the aim of latent optimisation is to traverse the latent space and find the optimal latent variable corresponding to the high-resolution result. The optimisation is formulated as a maximum likelihood problem minimising the consistency loss $L$.

\noindent \textbf{Multi-view loss function.} A typical CMR scan provides a stack of short-axis (SA) segmentations as the low-resolution segmentation $S_{LR}$. The loss function for short-axis view is defined as:
\begin{equation}
L_{SA}(z, d) = {CE}(M_{d}(\downarrow_s {G(z)}), S_{LR})
\label{Eq3}
\end{equation}
where $z$ and $d$ are the latent representation and inter-slice motion shift to be optimised and $CE$ denotes the cross-entropy loss, which describes the consistency between the degraded segmentation and input segmentation. In addition to the short-axis view, auxiliary views such as long-axis (LA) view segmentation $S_{LA}$ may be available. This provides an optional loss term $L_{LA}$:
\begin{equation}
L_{LA}(z, d) = {CE}(R\odot G(z), S_{LA})
\label{Eq4}
\end{equation}
where $R$ denotes the slicing operation that maps a volume to the LA view plane. The orientation of LA view is available from the DICOM header. The overall multi-view loss function is:
\begin{equation}
L(z, d) = L_{SA}(z, d) + \gamma \cdot L_{LA}(z, d)
\label{Eq5}
\end{equation}
where $\gamma$ is a hyper-parameter controlling the fidelity weight between SA and LA views. $\gamma$ is 0 if LA view is not available. To traverse the data manifold, the gradient of the loss function is backpropagated with regard to $z$ and $d$. We adopt a fully convolutional network architecture. The inter-slice motion shift and down-sampling are implemented as two differentiable layers (Figure \ref{Fig2}B).

\subsection{Implementation}
For the generative model $G$, we implemented 3D versions of $\beta$-VAE and deep convolutional generative adversarial networks (DCGAN). It turned out that $\beta$-VAE with 64 latent dimensions and appropriate regularisation ($\beta=0.001$) achieved the best performance. The models were implemented in PyTorch and trained using the Adam optimiser with a learning rate of 0.0001 and a batch size of 16. The architecture and loss functions as well as hyper-parameter studies are detailed in \textit{Supple. Material}. The weight $\gamma$ was set to 1 if long-axis view is available.The trained generative model and deployment code
are available at \url{https://github.com/shuowang26/SRHeart}.

For latent optimisation, we used the Adam optimiser with a learning rate of 0.0001. The initial values of $z$ and $d$ were set to 0. The iteration stops if the mean relative change of the loss is smaller than 5\%. The degradation process consists of two differentiable layers: the inter-slice motion layer and the down-sampling layer. Specifically, the motion layer was implemented by adding the slice-wise rigid motion onto the mesh grids, followed by bi-linear interpolation. The down-sampling layer was implemented with mesh grids scaled by the scaling factor and with bi-linear interpolation. 

For comparison, we implemented interpolation-based and supervised learning-based methods. For direct interpolation, we implemented the NN and SBI up-sampling \cite{raya1990shape}. For supervised learning, we implemented a 3D version of the enhanced deep residual network (EDSR) which is one of the state-of-the-art super-resolution CNNs \cite{anwar2020deep}.

\section{Experiments}
\subsection{Datasets}
\noindent \textbf{HR dataset:} This is a research cohort \cite{oktay2017anatomically} consisting of CMR images from 1,200 healthy volunteers, randomly split into three subsets: training (1,000), validation (100) and test (100). The spatial resolution is 1.25$\times$1.25$\times$2 mm which is regarded as `high-resolution' in this work. Cardiac structures of the left ventricle (LV), myocardium (MYO) and right ventricle (RV) were segmented using an atlas-based method, followed by manual correction and quality control.

\noindent \textbf{UK Biobank dataset:} This low-resolution dataset \cite{petersen2015uk} includes short-axis image stacks of 200 subjects, randomly selected from the whole UK Biobank study population ($\sim$100,000). The spatial resolution is 1.8$\times$1.8$\times$10 mm. The 4-chamber long-axis view is also available, which is a single-slice snapshot with a resolution of 1.8$\times$1.8 mm. The segmentations of LV, MYO and RV on both short-axis and long-axis views were performed automatically with a publicly available fully-convolutional network (FCN) \cite{bai2018automated}.

\subsection{Experimental design}

\noindent \textbf{Simulated degradation.}
We first simulate the degradation of high-resolution segmentations on the HR dataset using Equation (\ref{Eq1}). Technically, we set the down-sampling factor $s=[5, 1, 1]$ to simulate the low-resolution images acquired in the clinical routine. The direction and amplitude of inter-slice displacement of each slice is assumed independent and follow a Gaussian distribution with mean $\mu$=2.3 mm and standard variance $\sigma$=0.87 mm, which were fitted from previous studies on UK Biobank \cite{tarroni2020large}. To reflect the fact that respiratory motion could vary among scanning sites and populations, we also simulated scenarios with no inter-slice misalignment and with 4-times misalignment. Overall, we have curated three simulated HR-LR datasets corresponding to the different degradation patterns, named as the No-motion (down-sampling only), Normal-motion (down-sampling+normal respiratory motion), and Severe-motion (down-sampling+4$\times$normal respiratory motion) datasets.

\noindent \textbf{Real-world deployment.} \label{section54}
We evaluate the real-world performance on the UK Biobank dataset. As ground-truth high-resolution segmentations are not available for evaluation, we slice the super-resolved segmentation along the LA view and compare it to segmentation from LA view. We also investigate the benefits of multi-view loss by incorporating LA view segmentation into the framework. 

\section{Results and Discussion}

\subsection{Benchmark on simulated degradation }

\begin{figure}
\begin{center}
\includegraphics[width=\textwidth]{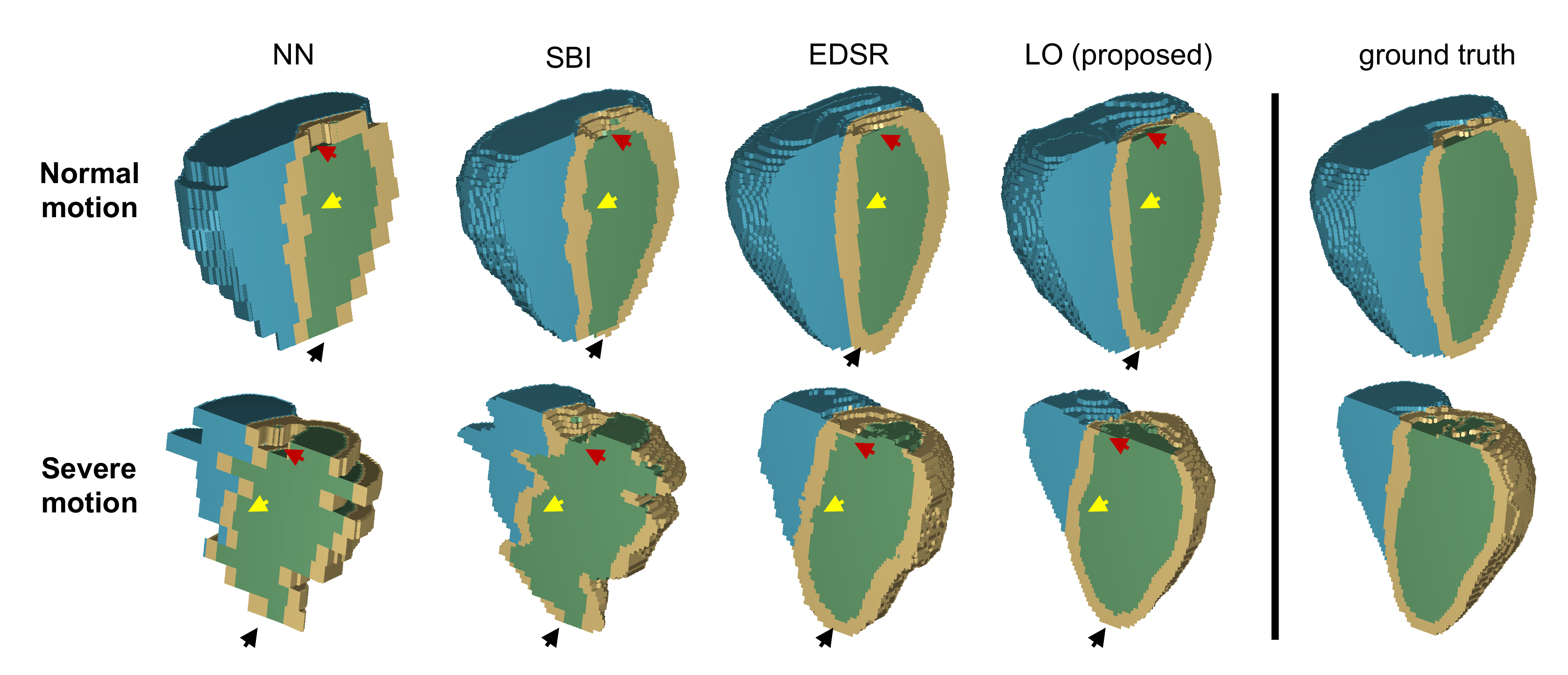}
\end{center}
   \caption{Illustration of model performance on the simulated Normal- and Severe-motion degradation. Enhancement near basal (red arrow), mid-ventricle (yellow arrow) and apical (black arrow) regions are compared.}
\label{Fig3}
\end{figure}

Examples of the super-resolution results from  NN, SBI, EDSR and the proposed LO model are demonstrated in Figure \ref{Fig3}. We used Dice coefficient as the metric for evaluating the segmentations. The quantitative results on the simulated datasets of different degradation patterns are reported in \textit{Supple. Material}. When respiratory motion is small (No-motion degradation), all models show good results. With Normal-motion degradation, the performance of all models dropped (\textit{Supple. Material}), especially for the most challenging MYO structure. In this scenario, our LO method showed comparable performance to EDSR and outperformed NN and SBI. Finally, if the respiration motion is high (Severe-motion), our method outperformed other models by a large margin (Table \ref{Tab1}). This suggests the robustness of the proposed framework. It is of clinical importance because patients with an acute disease could not follow the breath-holding instruction well, resulting in severe respiratory motions.

\begin{table}[htb]
\centering
\scriptsize
\begin{tabular}{|l|c|c|c|}
\hline
 & LV            & MYO           & RV            \\ \hline
NN     & 0.74$\pm$0.04 & 0.32$\pm$0.06 & 0.72$\pm$0.04 \\ \hline
SBI    & 0.75$\pm$0.03 & 0.31$\pm$0.07 & 0.74$\pm$0.04 \\ \hline
EDSR          & \textbf{0.89$\pm$0.02} & \textbf{0.70$\pm$0.07} & \textbf{0.89$\pm$0.03} \\ \hline
LO (proposed) & \textbf{0.92$\pm$0.02} & \textbf{0.75$\pm$0.08} & \textbf{0.91$\pm$0.02} \\ \hline

\end{tabular}
\caption{Comparison of the model performance under Severe-motion degradation. The Dice of different tissues (mean $\pm$ std) are evaluated between the ground truth and super-resolved 3D segmentations. Top 2 performance of each tissue are in bold.}
\label{Tab1}
\end{table}

\begin{table}[htb]
\centering
\scriptsize
\begin{tabular}{|l|l||l|l|l|}
\hline
   \multicolumn{2}{|c|}{} & \multicolumn{3}{c|}{Test domain}                                                           \\ \hline
& Training domain &
  \multicolumn{1}{c|}{\begin{tabular}[c]{@{}c@{}}No- \\ motion\end{tabular}} &
  \multicolumn{1}{c|}{\begin{tabular}[c]{@{}c@{}}Normal- \\ motion\end{tabular}} &
  \multicolumn{1}{c|}{\begin{tabular}[c]{@{}c@{}}Severe- \\ motion\end{tabular}} \\ \hline \hline
\multirow{3}{*}{EDSR} & No-motion   & \textbf{0.98$\pm$0.00} & 0.88$\pm$0.02 & 0.62$\pm$0.04 \\ \cline{2-5}
& Normal-motion & 0.91$\pm$0.01 & \textbf{0.93$\pm$0.01} & 0.70$\pm$0.05 \\ \cline{2-5}
& Severe-motion   & 0.68$\pm$0.03 & 0.75$\pm$0.03 & \textbf{0.83$\pm$0.04} \\ \hline \hline
\multicolumn{2}{|c|}{LO (proposed)}          & \textbf{0.95$\pm$0.01} & \textbf{0.92$\pm$0.01} & \textbf{0.86$\pm$0.04} \\ \hline 
\end{tabular}
\caption{Model generalisation under different simulated degradations. Mean Dice of three cardiac structures is evaluated. The first three rows demonstrate the domain-dependence of EDSR performance. Top 2 performance of each column are in bold.}
\label{Tab2}
\end{table}

We also demonstrate the advantage of our model over supervised learning-based EDSR method in terms of generalisation ability. As shown in Table \ref{Tab2}, the performance of EDSR dropped largely if degradation patterns of the training set and test set are different. Under these scenarios, the proposed model outperforms the EDSR models by a large margin ($>4\%$) with a better generalisation ability to unseen degradations.

\subsection{Real-world performance}

We deployed the models on the UK Biobank dataset. The EDSR model was trained on the simulated Normal-motion degradation. The Dice metric between the sliced super-resolved segmentation and the LA view segmentation (considered as ground truth) is reported in Table \ref{Tab3}. It shows that the LO method consistently outperformed the other models for all cardiac structures. In addition, incorporating the long-axis view information further improves the super-resolution performance.

\begin{table}[h]
\scriptsize 
\centering
\begin{tabular}{|l|l|l|l|}
\hline
 & LV       & MYO & RV \\ \hline \hline
NN    & 0.88 $\pm$ 0.07 & 0.57 $\pm$ 0.15 & 0.85 $\pm$ 0.12 \\ \hline
SBI   & 0.89 $\pm$ 0.07 & 0.58 $\pm$ 0.16 & 0.85 $\pm$ 0.12 \\ \hline
EDSR  & 0.90 $\pm$ 0.12 & 0.64 $\pm$ 0.12 & 0.87 $\pm$ 0.13 \\ \hline
LO (proposed)    & \textbf{0.91 $\pm$ 0.08} & \textbf{0.68 $\pm$ 0.12} & \textbf{0.87 $\pm$ 0.12} \\ \hline \hline
LO-multi-view & \textbf{0.92 $\pm$ 0.05} & \textbf{0.72 $\pm$ 0.10} & \textbf{0.89 $\pm$ 0.12} \\ \hline
\end{tabular}
\caption{Real-world performance on UK Biobank dataset evaluated on the 4-chamber long-axis view. LO-multi-view is the LO model incorporating the long-axis segmentation. Top performance of each column is in bold.}
\label{Tab3}
\end{table}

\section{Conclusion}
We proposed a latent optimisation framework for joint motion correction and super resolution for cardiac segmentation. It achieved high performance on both simulated and real-world datasets. It is a flexible framework, which is able to incorporate multi-view information and robust to domain shift.

\section*{Acknowledgements}
This research has been conducted using the UK Biobank Resource under Application Number 18545. The authors also acknowledge funding by EPSRC Programme (EP/P001009/1).

{
\bibliographystyle{splncs04}
\bibliography{Paper821.bbl}
}

\clearpage

\setcounter{figure}{0}
\setcounter{table}{0}
\renewcommand{\thefigure}{S\arabic{figure}}
\renewcommand{\thetable}{S\arabic{table}}

\setcounter{secnumdepth}{0}
\section{Supplementary Material: Joint Motion Correction and Super Resolution for Cardiac Segmentation via Latent Optimisation}

\begin{algorithm}
    \caption{Latent optimisation}\label{iter_scheme}
  \begin{algorithmic}[1]
    \REQUIRE Pre-trained generator $G:z \mapsto G(z)=S_{SR} \in \Sigma_{HR}$
    \INPUT A stack of low-resolution short-axis segmentations $S_{LR}$ and optional long-axis segmentations $S_{LA}$
    \OUTPUT Super-resolution result $S_{SR}$ and estimated inter-slice displacements $d$
    \STATE \textbf{Initialization} Latent variables $z=\mathbf{0}, d=\mathbf{0}$
    \WHILE{$L$ not converge}
        \STATE $L(z, d) = L_{SA}(z, d; S_{LR})+\gamma \cdot L_{LA}(z, d; S_{LA})$
        \STATE $z = z - \alpha \cdot \nabla_{z}L$ \COMMENT{learning rate $\alpha$}
        \STATE $d = d - \alpha \cdot \nabla_{d}L$
    \ENDWHILE
    \STATE $S_{SR}=G(z)$ 
  \end{algorithmic}
\end{algorithm}

\begin{figure}[h]
\begin{center}
\includegraphics[width=1\textwidth]{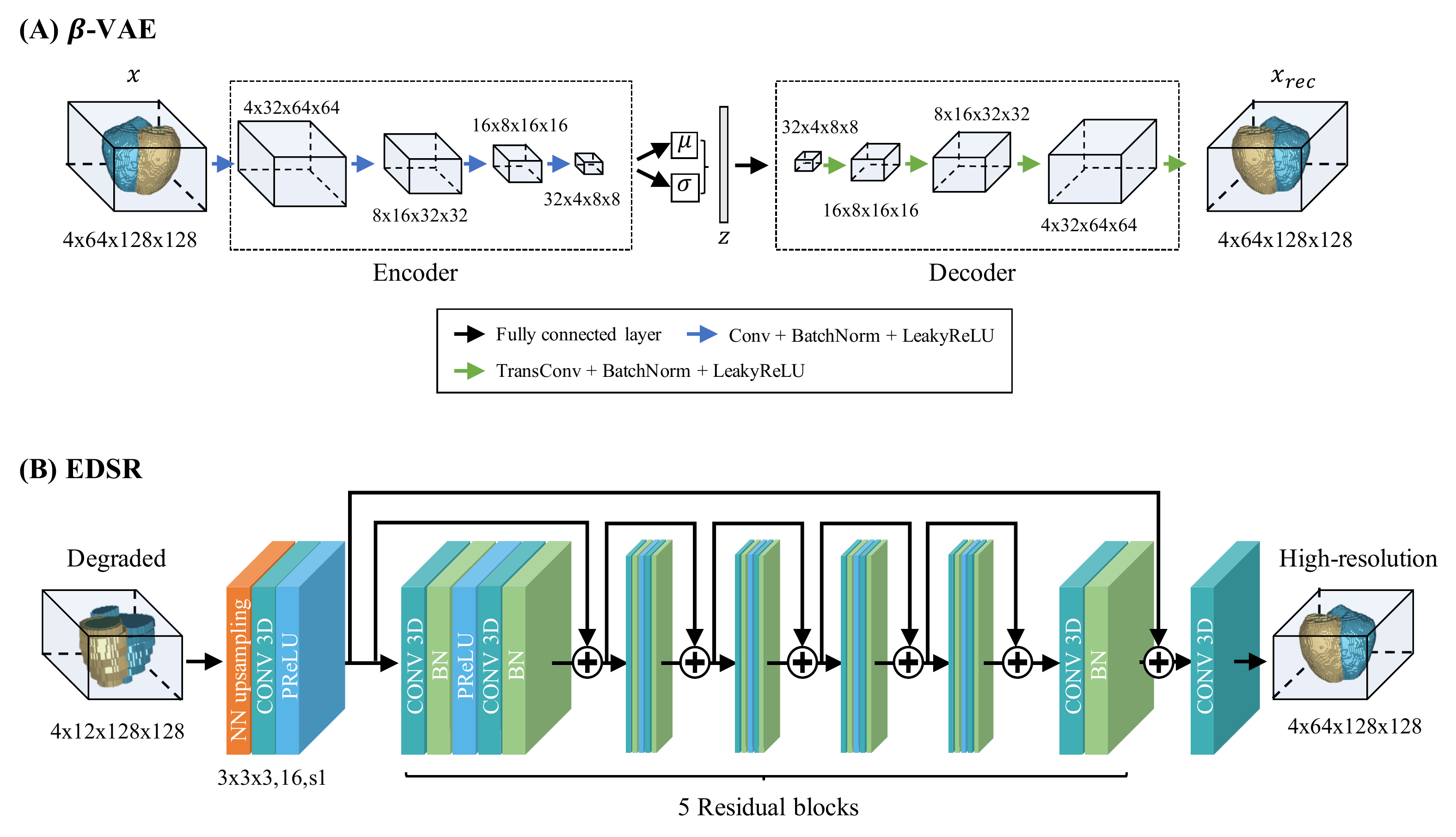}
\end{center}
\caption{Architectures of $\beta$-VAE and EDSR. (A) $\beta$-VAE was used as the generative model component of the proposed latent optimisation framework, described in Section 2.3 Implementation of the main text. The total loss for $\beta$-VAE is a combination of cross entropy loss and KL divergence loss weighted by $\beta$. (B) The 3D version of EDSR was re-implemented in this work to be compared with the proposed latent optimisation framework.} 

\label{FigS1}
\end{figure}

\begin{table}[htb]
\centering
\begin{tabular}{|c|c|c|c|c|c|}
\hline
Degradation levels                                   & Models & LV          & MYO         & RV          & MEAN        \\ \hline \hline
\multicolumn{1}{|c|}{\multirow{4}{*}{No motion}}     & NN     & 0.95 $\pm$ 0.01 & 0.86 $\pm$ 0.02 & 0.94 $\pm$ 0.01 & 0.92 $\pm$ 0.01 \\
\multicolumn{1}{|c|}{}                               & SBI    & 0.96 $\pm$ 0.01 & 0.87 $\pm$ 0.02 & 0.93 $\pm$ 0.01 & 0.92 $\pm$ 0.01 \\
\multicolumn{1}{|c|}{}                               & EDSR   & \textbf{0.99 $\pm$ 0.00} & \textbf{0.96 $\pm$ 0.01} & \textbf{0.98 $\pm$ 0.00} & \textbf{0.98 $\pm$ 0.00} \\
\multicolumn{1}{|c|}{}                               & LO (proposed)     & \textbf{0.97 $\pm$ 0.01} & \textbf{0.91 $\pm$ 0.01} & \textbf{0.96 $\pm$ 0.01} & \textbf{0.95 $\pm$ 0.01} \\ \hline \hline
\multicolumn{1}{|c|}{\multirow{4}{*}{Normal motion}} & NN     & 0.91 $\pm$ 0.01 & 0.72 $\pm$ 0.05 & 0.89 $\pm$ 0.01 & 0.84 $\pm$ 0.02 \\
\multicolumn{1}{|c|}{}                               & SBI    & 0.92 $\pm$ 0.01 & 0.73 $\pm$ 0.04 & 0.90 $\pm$ 0.01 & 0.85 $\pm$ 0.02 \\
\multicolumn{1}{|c|}{}                               & EDSR   & \textbf{0.96 $\pm$ 0.01} & \textbf{0.88 $\pm$ 0.02} & \textbf{0.96 $\pm$ 0.01} & \textbf{0.93 $\pm$ 0.01} \\
\multicolumn{1}{|c|}{}                               & LO (proposed)     & \textbf{0.96 $\pm$ 0.01} & \textbf{0.86 $\pm$ 0.02} & \textbf{0.95 $\pm$ 0.01} & \textbf{0.92 $\pm$ 0.01} \\ \hline \hline
\multicolumn{1}{|c|}{\multirow{4}{*}{Severe motion}} & NN     & 0.74 $\pm$ 0.04 & 0.32 $\pm$ 0.06 & 0.72 $\pm$ 0.04 & 0.59 $\pm$ 0.04 \\
\multicolumn{1}{|c|}{}                               & SBI    & 0.75 $\pm$ 0.03 & 0.31 $\pm$ 0.07 & 0.74 $\pm$ 0.04 & 0.60 $\pm$ 0.04 \\
\multicolumn{1}{|c|}{}                               & EDSR   & \textbf{0.89 $\pm$ 0.02} & \textbf{0.70 $\pm$ 0.07} & \textbf{0.89 $\pm$ 0.03} & \textbf{0.83 $\pm$ 0.04} \\
\multicolumn{1}{|c|}{}                               & LO (proposed)     & \textbf{0.92 $\pm$ 0.02} & \textbf{0.75 $\pm$ 0.08} & \textbf{0.91 $\pm$ 0.02} & \textbf{0.86 $\pm$ 0.04} \\ 
 \hline
\end{tabular}
\caption{Model performance under No-, Normal- and Severe-Motion degradation. For EDSR, models were trained with paired samples under one degradation level and evaluated on the same degradation level. Top 2 performance under each degradation level are in bold.}
\label{TabS3}
\end{table}

\end{document}